# A New Way for Determining Earthquake Magnitude
# Based on Electromagnetic Radiation

**Manana Kachakhidze [1],* Nino Kachakhidze-Murphy [1], Badri Khvitia,[2] Giorgi Ramishvili [3], Vaso Kukhianidze [3]**

1. Natural Hazard Scientific-Research Center, Georgian Technical University, Tbilisi 0175, Georgia
2. Sokhumi Institute of Physics and Technology, Tbilisi 0186, Georgia
3. School of Natural Sciences and Engineering, Ilia State University, Tbilisi 0162, Georgia

* Correspondence: kachakhidzem@gmail.com

## ABSTRACT

The article proposes the possibility of determining the magnitude of a future earthquake with high accuracy, based on the frequency data of VLF/LF electromagnetic emissions, as a diagnostic precursor that existed before an earthquake. Such magnitude will contribute to significantly improving the research of various fundamental issues in seismology in the future.

## 1. INTRODUCTION

Modern ground-based and satellite observations have revealed various anomalous changes in geophysical fields that attend the process of earthquake preparation in the lithosphere, as well as in the atmosphere and ionosphere. They are: changing of intensity of electrotelluric current in focal area; Perturbations of geomagnetic field in forms of irregular pulsations or regular short- period pulsations; Perturbations of atmospheric electric field; Irregular changing of characteristic parameters of the lower ionosphere (plasma frequency, electron content-ration, height of D layer, etc.); Irregular perturbations reaching the upper ionosphere, namely F2-layer, for 2-3 days before the earthquake; Increased intensity of electromagnetic emissions in upper ionosphere in several hours or tenths of minutes before earthquake; Lighting before earthquake; Infrared radiation; Total Electron Content (TEC) anomalies so on. Obviously, all the above mentioned phenomena are not observed before every earthquake and do not have a specific sequence [Gokhberg et al,1982; Uyeda et al., 2000; Hayakawa et.al., 2013; Biagi et.al., 2009; Pulinets et, al, 2011; Varotsos et.al., 2006; Parrot et.al., 2004-2010; Ouzounov et.al., 2006; Contadakis et.al., 2018].

It has been proven that among them there exists a geophysical field (VLF/LF) electromagnetic emissions that existed before the earthquakes), the anomalous changes of which accurately describe the geological processes of fault formation in the earthquake focus. Therefore, this field can be considered a diagnostic precursor. In such a case, it is no longer in dispute that it is possible to predict earthquakes based on it [Kachakhidze et. al., 2015, 2024]. It has been revealed that, based on this precursor field, it is possible to determine with very high accuracy the relationship between incoming earthquake fault length (L) and magnitude (M), which, in modern seismology, despite the very high level of development of this area of science, for the present remains one of the most important problems.

## 2. DISCUSSION

### 2.1. Some Information About Diagnostic Precursors of the Earthquakes.

As is known, the process of earthquake preparation proceeds in accordance with the well-known avalanche-like unstable geological model of fault formation, which step by step describes the occurrence of cracks of various sizes and the main process of forming the length of the fault in the focus of an impending earthquake. Namely, at the first stage, chaotic crack formation occurs in the earthquake preparation zone, when, as a result of the interaction of cracks, the average density of ruptures reaches a certain critical value in the entire volume or a significant part of it. As a result of the interaction of cracks, an avalanche stage of a given earthquake preparation sets in. It means that the first stage of earthquake preparation goes over to the second stage. The involvement of an increasing number of inhibited cracks in this process leads to the unification of cracks of a higher rank (larger size), which is associated with a rapid and sharp redistribution of the local stress field. An increase in the number and size of cracks, like an avalanche-like process, leads to a sharp increase in the rate of overall deformation, which causes a change in the integral physical characteristics of the medium. The increase in deformation is already accompanied by a drop in stress in the third stage. Due to the heterogeneity of the properties of the medium, the unstable deformation is compressed into a narrow zone, in which several relatively large cracks form. In this case, there is a general drop in average macro stresses over most of the volume. Due to this, the cracks stop developing and partially close. This process can be considered as the so-called main period of the "silence", which immediately begins after the end of the active stage of the avalanche process.

Already then, at the final stage, against the background of increasing tectonic tension, the fault of the earthquake is formed by ripping apart the cofferdam (links between cracks) in the focus of the upcoming earthquake.This process qualitatively is similar to the general process, and therefore it, by short-term and smaller changes in the amplitude of the strain rate, must be preceded. It happens that the destruction of one of the cofferdam may not be enough to rupture the entire main fault.

Due to this, there can be several such short-term changes in the deformation rate. In this case, as a result, these oscillations must precede large foreshocks, and the final one must be a direct foreshock or the main shock itself [Mjachkin et al., 1975].

It is widely believed in seismology that "One of the reasons making earthquake prediction so difficult is because we cannot monitor the stress and strain conditions of rocks a few kilometers below the ground surface where earthquakes are initiated' [Geller *et al.* 1997]. However, a geophysical field that makes it possible to trace step-by-step the geological process has now been discovered. This field is very low frequency/low frequency (VLF/LF) electromagnetic emissions existed before the earthquake. [Kachakhidze et.al., 2015, 2024].

It should be noted that research into the interaction between seismic activity and electromagnetic radiation, has intensified worldwide since the second half of the last century, as, alongside the theoretical research, it has become possible to conduct high-level laboratory and satellite experiments. Long-term observations have revealed special properties of VLF/LF electromagnetic radiation that manifest themselves both during the preparation of an earthquake and during the earthquake itself.

The networks to collect VLF/LF radio signals associated with seismic activity, since the end of the last century, in various seismically active regions and countries around the world have been established. One such network, based on the data of which our research was conducted, is the European Network for collecting VLF/LF radio signals (INFREP). [Biagi et al., 2009, 2013].

Before an earthquake, as mentioned above, many geophysical fields undergo anomalous changes, which is why all of these fields are referred to in the literature as precursors.

Our group has been studying anomalous changes in various geophysical fields before an earthquake for a long time, and concluded that, for further research in the right direction, it is necessary to classify geophysical fields into triggers, indicators, and precursor according to their generation mechanisms and role in the earthquake preparation process [Kachakhidze et al., 2023].

We think that for the observer, a precursor of an earthquake, or more precisely, a diagnostic precursor of the earthquake, is that geophysical field that is an exact manifestation of the geological stages of the avalanche-unstable model, and on the surface of the Earth, it describes the process of fault formation in the earthquake focus. A study of global scientific research allowed us to develop a model for the generation of electromagnetic radiation recorded before earthquakes, based on electrodynamics and the classical geological model of the avalanche-like process of fault formation. This is the model of electromagnetic radiation of the lithospheric origin that exists before and during an earthquake. The seismogenic area is considered to be an oscillatory-distributed system. This model simplifies physical analyses of the nonlinear effects and qualitatively explains the mechanisms that generate very low frequency electromagnetic waves in the period prior to an earthquake.

Using the VLF/LF electromagnetic radiation that exists in the period preceding the earthquake, the authors obtained formula (1), which, for the first time, analytically links the frequency of the observed electromagnetic radiation and the linear size (length of the fault at the source) of the earthquake:

$$\omega = k\frac{c}{l} \qquad (1)$$

where *c* is the velocity of light, $l$ is the length of the rupture, *k* is the characteristic coefficient of geological medium and it approximately equals to 1 [Kachakhidze et al., 2015]. It should be emphasized that the very high accuracy of measuring electromagnetic radiation in the VLF/LF ranges also makes it possible to determine the length of small cracks with high accuracy. For instance, before the L’Aquila catastrophic earthquake (6 April 2009), besides the other electromagnetic anomalies, 41 MHz ($\omega_1$) and about 54 MHz ($\omega_2$) radiation were also recorded [Eftaxias et al., 2002)] According to our study (formula 1), the corresponding length of the first emitting body (which radiates $\omega_1$ frequency) would be $l_1$ = 0.007 km, and the corresponding length of the second emitting body (which emitted $\omega_2$ frequency) would be $l_2$ = 0.006 km. This means that electromagnetic radiation can detect even the slightest changes in the length of the microcracks.

Studies have shown that formula (1) is in good agreement with real earthquake data [Kachakhidze et al., 2019, 2024].

To relate magnitude and rupture length $l$, we used two formulas: for $M_S$ (2) [Ulomov, 1993] and for $M_W$ (3) [Wells, et al.,1994] for M ≥5.0 earthquakes:

$$lgl = 0.6M_s - 2.5 \quad (2),$$
$$M_w = 4.38 + 1.49 * \log l \quad (3)$$

It should be noted that, unlike satellite-based studies, work relying on data from ground-based networks make it possible to record the electromagnetic field at the very moment an earthquake occurs. [Kachakhidze et al., 2019, 2022, 2023, 2024].

## 2.2. Possibility of Increasing the Accuracy of Earthquake Magnitude Measurements Based on Modern Research

Accurate measurement of earthquake magnitude remains one of the main tasks of seismology to this day. The famous scientist C. F. Richter wrote that, "the local magnitude ($M_L$) requires only a source-receiver distance and a measurement of the peak amplitude of a filtered seismogram". [Richter, 1935]. "Since then, different studies have developed local magnitude formulas that account for the local attenuation properties of a particular region. Despite such widespread use of $M_L$, the consistency and accuracy of $M_L$ are challenged by large variations in estimates from station to station, influences of radiation pattern and rupture directivity, and spatially variable attenuation structure. In addition, highlights that $M_L$ might not be physically meaningful for small events, as the systematic and random errors in its determination are likely greater than its variability relative to moment magnitude estimates" [Castro et al, 2026].

"Conventional magnitude scales, including mb, Ms, and ML, encounter saturation issues when the earthquake's rupture dimension exceeds the wavelength of the seismic waves used for magnitude determination. For large earthquakes, rupture dimensions often exceed the wavelengths of seismic waves commonly employed for magnitude estimation" [Das et al, 2024].

Almost every famous seismologist in the world has studied and continues to study this issue, but the problem has still not been solved.

Based on the above, we propose a new way for measuring earthquake magnitude that can be used in any region for the entire magnitude spectrum, covering not only small and medium, but also large and very large earthquakes, which is another additional problem in seismology.

In the tables I and II are given data of the fault length, corresponding to the magnitude, measured to tenths of accuracy, as accepted in seismology and of the VLF/LF electromagnetic emissions for earthquakes of magnitudes M≥5, calculated using formulas (1), (2), and (3), and data of the relevant energies calculated by formulas (4) and (5):

$$lg_{10} E = 4.8 + 1.5\, M_S \quad (4)$$
$$log_{10} E = 5.24 + 1.44.\, M_w \quad (5)$$

It should be noted that the lengths of the faults that occurred in the focus of the earthquake are given with an accuracy even of cm, because calculations performed with VLF/LF electromagnetic emissions allow us to measure the existing fault with such accuracy.

**Table I** for $M_S$ magnitude

| Ms | $l$ (km) | Frequency (Hz) | Energy E (joule) |
|---|---|---|---|
| 5 | 3.16228- 3.62577 | 94868-82741 | 2.00E+12-2.81E+12 |
| 5.1 | 3.63078-4.16294 | 82626-72064 | 2.82E+12-3.97E+12 |
| 5.2 | 4.16869-4.77969 | 71965-62765 | 3.98E+12-5.60E+12 |
| 5.3 | 4.7863-5.48782 | 62678-54666 | 5.62E+12-7.92E+12 |
| 5.4 | 5.49541-6.30086 | 54591-47612 | 7.94E+12-1.12E+13 |
| 5.5 | 6.30957-7.23436 | 47546-41468 | 1.12E+13-1.58E+13 |
| 5.6 | 7.24436-8.30615 | 41411-36117 | 1.58E+13-2.23E+13 |
| 5.7 | 8.31764-9.53674 | 36067-31457 | 2.24E+13-3.15E+13 |
| 5.8 | 9.54993-10.9496 | 31413-27398 | 3.16E+13-4.45E+13 |
| 5.9 | 10.9648-12.5719 | 27360-23862 | 4.47E+13-6.29E+13 |
| 6 | 12.5893-14.4344 | 23829-20783 | 6.31E+13-8.88E+13 |
| 6.1 | 14.4544-16.573 | 20754-18101 | 8.91E+13-1.25E+14 |
| 6.2 | 16.5959-19.0283 | 18076-15766 | 1.26E+14-1.77E+14 |
| 6.3 | 19.0546-21.8474 | 15744-13731 | 1.78E+14-2.50E+14 |
| 6.4 | 21.8776-25.0842 | 13712-11959 | 2.51E+14-3.54E+14 |
| 6.5 | 25.1189-28.8005 | 11943-10416 | 3.55E+14-4.99E+14 |
| 6.6 | 28.8403-33.0674 | 10402-9072 | 5.01E+14-7.06E+14 |
| 6.7 | 33.1131-37.9665 | 9059-7901 | 7.08E+14-9.97E+14 |
| 6.8 | 38.0189-43.5913 | 7890-6882 | 1.00E+15-1.41E+15 |
| 6.9 | 43.6516-50.0495 | 6872-5994 | 1.41E+15-1.99E+15 |
| 7 | 50.1187-57.4645 | 5985-5220 | 2.00E+15-2.81E+15 |
| 7.1 | 57.544-65.9781 | 5213-4547 | 2.82E+15-3.97E+15 |
| 7.2 | 66.0693-75.753 | 4540-3960 | 3.98E+15-5.60E+15 |
| 7.3 | 75.8578-86.9761 | 3954-3449 | 5.62E+15-7.92E+15 |
| 7.4 | 87.0964-99.8619 | 3444-3004 | 7.94E+15-1.12E+16 |
| 7.5 | 100-114.657 | 3000-2616 | 1.12E+16-1.58E+16 |
| 7.6 | 114.815-131.644 | 2612-2278 | 1.58E+16-2.23E+16 |
| 7.7 | 131.826-151.147 | 2275-1984 | 2.24E+16-3.15E+16 |
| 7.8 | 151.356-173.54 | 1982-1728 | 3.16E+16-4.45E+16 |
| 7.9 | 173.78-199.251 | 1726-1505 | 4.47E+16-6.29E+16 |
| 8 | 199.526-228.77 | 1503-1311 | 6.31E+16-8.88E+16 |
| 8.1 | 229.087-262.664 | 1309-1142 | 8.91E+16-1.25E+17 |
| 8.2 | 263.027-301.578 | 1140-994 | 1.26E+17-1.77E+17 |
| 8.3 | 301.995-346.258 | 993-866 | 1.78E+17-2.50E+17 |
| 8.4 | 346.737-397.558 | 865-754 | 2.51E+17-3.54E+17 |
| 8.5 | 398.107-456.457 | 753-657 | 3.55E+17-4.99E+17 |
| 8.6 | 457.088-524.083 | 656-572 | 5.01E+17-7.06E+17 |
| 8.7 | 524.807-601.728 | 571-498 | 7.08E+17-9.97E+17 |
| 8.8 | 602.56-690.876 | 497-434 | 1.00E+18-1.41E+18 |
| 8.9 | 691.831-793.232 | 433-377 | 1.41E+18-1.99E+18 |
| 9 | 794.328-910.752 | 377-329 | 2.00E+18-2.81E+18 |

**Table II** for $\boldsymbol{M_w}$ magnitude

| Mw | $l$ (km) | Frequency (Hz) | Energy E (joule) |
|---|---|---|---|
| 5 | 2.60679-3.03773 | 115083-98757 | 2.75E+12-3.82E+12 |
| 5.1 | 3.04243-3.54539 | 98605-84616 | 3.84E+12-5.33E+12 |
| 5.2 | 3.55087-4.13788 | 84486-72500 | 5.35E+12-7.42E+12 |
| 5.3 | 4.14428-4.82939 | 72388-62119 | 7.45E+12-1.03E+13 |
| 5.4 | 4.83686-5.63646 | 62023-53224 | 1.04E+13-1.44E+13 |
| 5.5 | 5.64518-6.57840 | 53142-45603 | 1.45E+13-2.01E+13 |
| 5.6 | 6.58858-7.67776 | 45533-39073 | 2.01E+13-2.80E+13 |
| 5.7 | 7.68964-8.96084 | 39013-33479 | 2.81E+13-3.90E+13 |
| 5.8 | 8.97470-10.45835 | 33427-28685 | 3.91E+13-5.43E+13 |
| 5.9 | 10.4745-12.20611 | 28640-24577 | 5.45E+13-7.56E+13 |
| 6 | 12.22498-14.24595 | 24539-21058 | 7.59E+13-1.05E+14 |
| 6.1 | 14.26798-16.62667 | 21026-18043 | 1.06E+14-1.47E+14 |
| 6.2 | 16.65239-19.40526 | 18015-15459 | 1.47E+14-2.04E+14 |
| 6.3 | 19.43527-22.64819 | 15435-13246 | 2.05E+14-2.85E+14 |
| 6.4 | 22.68322-26.43307 | 13225-11349 | 2.86E+14-3.97E+14 |
| 6.5 | 26.47395-30.85047 | 11331-9724 | 3.98E+14-5.53E+14 |
| 6.6 | 30.89818-36.00608 | 9709-8331 | 5.55E+14-7.70E+14 |
| 6.7 | 36.06177-42.02328 | 8319-7138 | 7.73E+14-1.07E+15 |
| 6.8 | 42.08827-49.04605 | 7127-6116 | 1.08E+15-1.49E+15 |
| 6.9 | 49.1219-57.24244 | 6107-5240 | 1.50E+15-2.08E+15 |
| 7 | 57.33097-66.80858 | 5232-4490 | 2.09E+15-2.90E+15 |
| 7.1 | 66.9119-77.97337 | 4483-3847 | 2.91E+15-4.04E+15 |
| 7.2 | 78.09396-91.00398 | 3841-3296 | 4.06E+15-5.63E+15 |
| 7.3 | 91.14473-106.2122 | 3291-2824 | 5.65E+15-7.84E+15 |
| 7.4 | 106.3765-123.962 | 2820-2420 | 7.87E+15-1.09E+16 |
| 7.5 | 124.1537-144.6781 | 2416-2073 | 1.10E+16-1.52E+16 |

| 7.6 | 144.9018-168.8561 | 2070-1776 | 1.53E+16-2.12E+16 |
|---|---|---|---|
| 7.7 | 169.1173-197.0747 | 1773-1522 | 2.13E+16-2.96E+16 |
| 7.8 | 197.3795-230.0091 | 1519-1304 | 2.96E+16-4.12E+16 |
| 7.9 | 230.3648-268.4473 | 1302-1117 | 4.13E+16-5.74E+16 |
| 8 | 268.8625-313.3092 | 1115-957 | 5.75E+16-7.99E+16 |
| 8.1 | 313.7938-365.6682 | 956-820 | 8.02E+16-1.11E+17 |
| 8.2 | 366.2338-426.7773 | 819-702 | 1.12E+17-1.55E+17 |
| 8.3 | 427.4374-498.0987 | 701-602 | 1.56E+17-2.16E+17 |
| 8.4 | 498.8691-581.3391 | 601-516 | 2.17E+17-3.01E+17 |
| 8.5 | 582.2382-678.4903 | 515-442 | 3.02E+17-4.19E+17 |
| 8.6 | 679.5396-791.877 | 441-378 | 4.21E+17-5.84E+17 |
| 8.7 | 793.1017-924.2124 | 378-324 | 5.86E+17-8.14E+17 |
| 8.8 | 925.6418-1078.663 | 324-278 | 8.17E+17-1.13E+18 |
| 8.9 | 1080.332-1258.925 | 277-238 | 1.14E+18-1.58E+18 |
| 9 | 1260.872-1469.312 | 237-204 | 1.58E+18-2.20E+18 |

The fact is that, based on studies of VLF/VLF electromagnetic radiation, that existed before the earthquake, it was established that to each value of $M_s$ and $M_w$ magnitudes, measured with an accuracy of one tenth, corresponds a certain diapason of rupture length and a certain diapason of earthquake energy.

To overcome this difficulty, we set ourselves the task: with what accuracy should the magnitude of a future earthquake be measured so that one specific magnitude value corresponds to one specific rupture length value and one specific earthquake energy value?

This paper proposes a method to determine the rupture length (i.e., magnitude) of a future earthquake with higher accuracy, and thus provides a completely different way to solve the problem of accurately determining earthquake energy.

In particular, it is proposed to measure the frequency of VLF/LF electromagnetic radiation before an earthquake to within 0.1 Hz, which will make it possible the earthquake's magnitude has already been estimated to within 0.001 magnitudes. In turn, a specific, unique magnitude value, measured to within 0.001, corresponds to a unique rupture length value, measured to a precision of one meter, and determines a unique, specific value for the earthquake energy.

As an example, the results of calculations for earthquakes $7.4 \leq M \leq 7.499$ in Tables III and IV are given. This order of columns in the tables is conditioned by the nature of the set task at hand.

When monitoring the earthquake preparation process, the frequency of the VLF/LF electromagnetic emissions is necessary to be measured to within 0.1, which ensures the uniqueness of all other parameters.

Tables III and IV present data on the frequency of electromagnetic radiation (measured with an accuracy of 0.1 Hz ), the length of the rupture with an accuracy of one meter, and the energy (measured in joules) for cases of earthquakes of (**Ms** and **Mw**) 7.4 ≤ M ≤ 7.499. Magnitude measured with an accuracy of 0.001.

**Table III** for $M_s$

| $M_s$ | Length (km) | Frequency (Hz) | Energy (joule) |
|---|---|---|---|
| 7.4 | 87.096 | 3444.5 | 7.943E+15 |
| 7.401 | 87.217 | 3439.7 | 7.971E+15 |
| 7.402 | 87.337 | 3435.0 | 7.998E+15 |
| 7.403 | 87.458 | 3430.2 | 8.026E+15 |
| 7.404 | 87.579 | 3425.5 | 8.054E+15 |
| 7.405 | 87.7 | 3420.7 | 8.082E+15 |
| 7.406 | 87.821 | 3416.0 | 8.110E+15 |
| 7.407 | 87.943 | 3411.3 | 8.138E+15 |
| 7.408 | 88.064 | 3406.6 | 8.166E+15 |
| 7.409 | 88.186 | 3401.9 | 8.194E+15 |
| 7.41 | 88.308 | 3397.2 | 8.222E+15 |
| 7.411 | 88.43 | 3392.5 | 8.251E+15 |
| 7.412 | 88.552 | 3387.8 | 8.279E+15 |
| 7.413 | 88.675 | 3383.2 | 8.308E+15 |
| 7.414 | 88.797 | 3378.5 | 8.337E+15 |
| 7.415 | 88.92 | 3373.8 | 8.366E+15 |
| 7.416 | 89.043 | 3369.2 | 8.395E+15 |
| 7.417 | 89.166 | 3364.5 | 8.424E+15 |
| 7.418 | 89.289 | 3359.9 | 8.453E+15 |
| 7.419 | 89.413 | 3355.2 | 8.482E+15 |
| 7.42 | 89.536 | 3350.6 | 8.511E+15 |
| 7.421 | 89.66 | 3346.0 | 8.541E+15 |
| 7.422 | 89.784 | 3341.3 | 8.570E+15 |
| 7.423 | 89.908 | 3336.7 | 8.600E+15 |
| 7.424 | 90.033 | 3332.1 | 8.630E+15 |
| 7.425 | 90.157 | 3327.5 | 8.660E+15 |
| 7.426 | 90.282 | 3322.9 | 8.690E+15 |
| 7.427 | 90.407 | 3318.3 | 8.720E+15 |
| 7.428 | 90.532 | 3313.8 | 8.750E+15 |
| 7.429 | 90.657 | 3309.2 | 8.780E+15 |
| 7.43 | 90.782 | 3304.6 | 8.810E+15 |
| 7.431 | 90.908 | 3300.1 | 8.841E+15 |
| 7.432 | 91.033 | 3295.5 | 8.872E+15 |
| 7.433 | 91.159 | 3290.9 | 8.902E+15 |
| 7.434 | 91.285 | 3286.4 | 8.933E+15 |
| 7.435 | 91.411 | 3281.9 | 8.964E+15 |
| 7.436 | 91.538 | 3277.3 | 8.995E+15 |
| 7.437 | 91.664 | 3272.8 | 9.026E+15 |
| 7.438 | 91.791 | 3268.3 | 9.057E+15 |
| 7.439 | 91.918 | 3263.8 | 9.089E+15 |
| 7.44 | 92.045 | 3259.3 | 9.120E+15 |
| 7.441 | 92.172 | 3254.8 | 9.152E+15 |
| 7.442 | 92.3 | 3250.3 | 9.183E+15 |
| 7.443 | 92.427 | 3245.8 | 9.215E+15 |
| 7.444 | 92.555 | 3241.3 | 9.247E+15 |
| 7.445 | 92.683 | 3236.8 | 9.279E+15 |
| 7.446 | 92.811 | 3232.4 | 9.311E+15 |
| 7.447 | 92.939 | 3227.9 | 9.343E+15 |
| 7.448 | 93.068 | 3223.5 | 9.376E+15 |
| 7.449 | 93.197 | 3219.0 | 9.408E+15 |
| 7.45 | 93.325 | 3214.6 | 9.441E+15 |
| 7.451 | 93.454 | 3210.1 | 9.473E+15 |
| 7.452 | 93.584 | 3205.7 | 9.506E+15 |
| 7.453 | 93.713 | 3201.3 | 9.539E+15 |
| 7.454 | 93.843 | 3196.8 | 9.572E+15 |
| 7.455 | 93.972 | 3192.4 | 9.605E+15 |
| 7.456 | 94.102 | 3188.0 | 9.638E+15 |
| 7.457 | 94.232 | 3183.6 | 9.672E+15 |
| 7.458 | 94.363 | 3179.2 | 9.705E+15 |
| 7.459 | 94.493 | 3174.8 | 9.739E+15 |
| 7.46 | 94.624 | 3170.5 | 9.772E+15 |
| 7.461 | 94.755 | 3166.1 | 9.806E+15 |
| 7.462 | 94.886 | 3161.7 | 9.840E+15 |
| 7.463 | 95.017 | 3157.3 | 9.874E+15 |
| 7.464 | 95.148 | 3153.0 | 9.908E+15 |
| 7.465 | 95.28 | 3148.6 | 9.943E+15 |
| 7.466 | 95.411 | 3144.3 | 9.977E+15 |
| 7.467 | 95.543 | 3139.9 | 1.001E+16 |
| 7.468 | 95.675 | 3135.6 | 1.005E+16 |
| 7.469 | 95.808 | 3131.3 | 1.008E+16 |
| 7.47 | 95.94 | 3127.0 | 1.012E+16 |
| 7.471 | 96.073 | 3122.6 | 1.015E+16 |
| 7.472 | 96.206 | 3118.3 | 1.019E+16 |
| 7.473 | 96.339 | 3114.0 | 1.022E+16 |

| 7.474 | 96.472 | 3109.7 | 1.026E+16 |
|---|---|---|---|
| 7.475 | 96.605 | 3105.4 | 1.029E+16 |
| 7.476 | 96.739 | 3101.1 | 1.033E+16 |
| 7.477 | 96.872 | 3096.9 | 1.036E+16 |
| 7.478 | 97.006 | 3092.6 | 1.040E+16 |
| 7.479 | 97.14 | 3088.3 | 1.044E+16 |
| 7.48 | 97.275 | 3084.0 | 1.047E+16 |
| 7.481 | 97.409 | 3079.8 | 1.051E+16 |
| 7.482 | 97.544 | 3075.5 | 1.054E+16 |
| 7.483 | 97.679 | 3071.3 | 1.058E+16 |
| 7.484 | 97.814 | 3067.1 | 1.062E+16 |
| 7.485 | 97.949 | 3062.8 | 1.065E+16 |
| 7.486 | 98.084 | 3058.6 | 1.069E+16 |
| 7.487 | 98.22 | 3054.4 | 1.073E+16 |
| 7.488 | 98.356 | 3050.2 | 1.076E+16 |
| 7.489 | 98.492 | 3045.9 | 1.080E+16 |
| 7.49 | 98.628 | 3041.7 | 1.084E+16 |
| 7.491 | 98.764 | 3037.5 | 1.088E+16 |
| 7.492 | 98.901 | 3033.3 | 1.091E+16 |
| 7.493 | 99.038 | 3029.2 | 1.095E+16 |
| 7.494 | 99.174 | 3025.0 | 1.099E+16 |
| 7.495 | 99.312 | 3020.8 | 1.103E+16 |
| 7.496 | 99.449 | 3016.6 | 1.107E+16 |
| 7.497 | 99.586 | 3012.5 | 1.110E+16 |
| 7.498 | 99.724 | 3008.3 | 1.114E+16 |
| 7.499 | 99.862 | 3004.1 | 1.118E+16 |

**Table IV** for $M_w$

| $M_w$ | Length (km) | Frequency (Hz) | Energy (joule) |
|---|---|---|---|
| 7.4 | 106.376 | 2820.2 | 7.870E+15 |
| 7.401 | 106.541 | 2815.8 | 7.897E+15 |
| 7.402 | 106.706 | 2811.5 | 7.923E+15 |
| 7.403 | 106.871 | 2807.1 | 7.949E+15 |
| 7.404 | 107.036 | 2802.8 | 7.976E+15 |
| 7.405 | 107.202 | 2798.5 | 8.002E+15 |
| 7.406 | 107.367 | 2794.1 | 8.029E+15 |
| 7.407 | 107.533 | 2789.8 | 8.055E+15 |
| 7.408 | 107.7 | 2785.5 | 8.082E+15 |
| 7.409 | 107.866 | 2781.2 | 8.109E+15 |
| 7.41 | 108.033 | 2776.9 | 8.136E+15 |
| 7.411 | 108.2 | 2772.6 | 8.163E+15 |
| 7.412 | 108.368 | 2768.4 | 8.190E+15 |
| 7.413 | 108.535 | 2764.1 | 8.217E+15 |
| 7.414 | 108.703 | 2759.8 | 8.244E+15 |
| 7.415 | 108.871 | 2755.6 | 8.272E+15 |
| 7.416 | 109.04 | 2751.3 | 8.299E+15 |
| 7.417 | 109.208 | 2747.0 | 8.327E+15 |
| 7.418 | 109.377 | 2742.8 | 8.354E+15 |
| 7.419 | 109.546 | 2738.6 | 8.382E+15 |
| 7.42 | 109.716 | 2734.3 | 8.410E+15 |
| 7.421 | 109.885 | 2730.1 | 8.438E+15 |
| 7.422 | 110.055 | 2725.9 | 8.466E+15 |
| 7.423 | 110.225 | 2721.7 | 8.494E+15 |
| 7.424 | 110.396 | 2717.5 | 8.522E+15 |
| 7.425 | 110.567 | 2713.3 | 8.551E+15 |
| 7.426 | 110.738 | 2709.1 | 8.579E+15 |
| 7.427 | 110.909 | 2704.9 | 8.608E+15 |
| 7.428 | 111.08 | 2700.7 | 8.636E+15 |
| 7.429 | 111.252 | 2696.6 | 8.665E+15 |
| 7.43 | 111.424 | 2692.4 | 8.694E+15 |
| 7.431 | 111.597 | 2688.3 | 8.722E+15 |
| 7.432 | 111.769 | 2684.1 | 8.751E+15 |
| 7.433 | 111.942 | 2680.0 | 8.781E+15 |
| 7.434 | 112.115 | 2675.8 | 8.810E+15 |
| 7.435 | 112.289 | 2671.7 | 8.839E+15 |
| 7.436 | 112.462 | 2667.6 | 8.868E+15 |
| 7.437 | 112.636 | 2663.4 | 8.898E+15 |
| 7.438 | 112.81 | 2659.3 | 8.927E+15 |
| 7.439 | 112.985 | 2655.2 | 8.957E+15 |
| 7.44 | 113.16 | 2651.1 | 8.987E+15 |
| 7.441 | 113.335 | 2647.0 | 9.017E+15 |
| 7.442 | 113.51 | 2642.9 | 9.046E+15 |
| 7.443 | 113.685 | 2638.9 | 9.077E+15 |
| 7.444 | 113.861 | 2634.8 | 9.107E+15 |
| 7.445 | 114.037 | 2630.7 | 9.137E+15 |
| 7.446 | 114.214 | 2626.7 | 9.167E+15 |
| 7.447 | 114.39 | 2622.6 | 9.198E+15 |

| 7.448 | 114.567 | 2618.5 | 9.228E+15 |
|---|---|---|---|
| 7.449 | 114.744 | 2614.5 | 9.259E+15 |
| 7.45 | 114.922 | 2610.5 | 9.290E+15 |
| 7.451 | 115.1 | 2606.4 | 9.321E+15 |
| 7.452 | 115.278 | 2602.4 | 9.351E+15 |
| 7.453 | 115.456 | 2598.4 | 9.383E+15 |
| 7.454 | 115.634 | 2594.4 | 9.414E+15 |
| 7.455 | 115.813 | 2590.4 | 9.445E+15 |
| 7.456 | 115.992 | 2586.4 | 9.476E+15 |
| 7.457 | 116.172 | 2582.4 | 9.508E+15 |
| 7.458 | 116.351 | 2578.4 | 9.539E+15 |
| 7.459 | 116.531 | 2574.4 | 9.571E+15 |
| 7.46 | 116.712 | 2570.4 | 9.603E+15 |
| 7.461 | 116.892 | 2566.5 | 9.635E+15 |
| 7.462 | 117.073 | 2562.5 | 9.667E+15 |
| 7.463 | 117.254 | 2558.5 | 9.699E+15 |
| 7.464 | 117.435 | 2554.6 | 9.731E+15 |
| 7.465 | 117.617 | 2550.7 | 9.763E+15 |
| 7.466 | 117.799 | 2546.7 | 9.796E+15 |
| 7.467 | 117.981 | 2542.8 | 9.828E+15 |
| 7.468 | 118.163 | 2538.9 | 9.861E+15 |
| 7.469 | 118.346 | 2534.9 | 9.894E+15 |
| 7.47 | 118.529 | 2531.0 | 9.927E+15 |
| 7.471 | 118.713 | 2527.1 | 9.960E+15 |
| 7.472 | 118.896 | 2523.2 | 9.993E+15 |
| 7.473 | 119.08 | 2519.3 | 1.003E+16 |

| 7.474 | 119.264 | 2515.4 | 1.006E+16 |
|---|---|---|---|
| 7.475 | 119.449 | 2511.5 | 1.009E+16 |
| 7.476 | 119.633 | 2507.7 | 1.013E+16 |
| 7.477 | 119.818 | 2503.8 | 1.016E+16 |
| 7.478 | 120.004 | 2499.9 | 1.019E+16 |
| 7.479 | 120.189 | 2496.1 | 1.023E+16 |
| 7.48 | 120.375 | 2492.2 | 1.026E+16 |
| 7.481 | 120.561 | 2488.4 | 1.030E+16 |
| 7.482 | 120.748 | 2484.5 | 1.033E+16 |
| 7.483 | 120.935 | 2480.7 | 1.036E+16 |
| 7.484 | 121.122 | 2476.9 | 1.040E+16 |
| 7.485 | 121.309 | 2473.0 | 1.043E+16 |
| 7.486 | 121.496 | 2469.2 | 1.047E+16 |
| 7.487 | 121.684 | 2465.4 | 1.050E+16 |
| 7.488 | 121.873 | 2461.6 | 1.054E+16 |
| 7.489 | 122.061 | 2457.8 | 1.057E+16 |
| 7.49 | 122.25 | 2454.0 | 1.061E+16 |
| 7.491 | 122.439 | 2450.2 | 1.064E+16 |
| 7.492 | 122.628 | 2446.4 | 1.068E+16 |
| 7.493 | 122.818 | 2442.6 | 1.071E+16 |
| 7.494 | 123.008 | 2438.9 | 1.075E+16 |
| 7.495 | 123.198 | 2435.1 | 1.078E+16 |
| 7.496 | 123.389 | 2431.3 | 1.082E+16 |
| 7.497 | 123.579 | 2427.6 | 1.086E+16 |
| 7.498 | 123.771 | 2423.8 | 1.089E+16 |
| 7.499 | 123.962 | 2420.1 | 1.093E+16 |

From the tables III and IV it is clear that one value of the frequency of electromagnetic radiation corresponds to only one value of the magnitude, length of the rupture, and energy of the earthquake.

We have conducted similar calculations for earthquakes of magnitude $1 \leq M \leq 9$, but due to the large volume of data, we could not include the corresponding tables in the article.

The proposed method will allow for the accurate solution of some fundamental problems of seismology in the future.

As we noted above, in our model, the seismogenic area is considered to be an oscillatory-distributed system, which is characterized by the following features:

1. The segment of earth crust, where incoming earthquake focus is to be formed from the very starting moment of earthquake preparation, belongs to the system, which suffers specific type oscillations: the process of energy accumulation is in progress in the system. But as a result of foreshocks, the main shock, and aftershocks, the accumulated energy is released. Taking this into account, this system is the oscillatory system. [Migulin et al., 1978]
2. In the seismogenic area, the mass, elasticity (mechanical systems), capacity, and inductance (electric systems) elements are uniformly and uninterruptedly spread in the whole volume of the system.

In addition, due to piezoelectric, piezomagnetic, electrochemical, and other effects, every tiny element in the earthquake preparation zone has its capacitance and inductance. Therefore seismogenic zone can be considered a distributed system simultaneously. The relationship between the arrangement of elements (or groups of elements) plays a significant role in terms of the functioning of the system, since this system is constantly subject to tectonic stress, which changes the physical and chemical properties of the environment.

As a rule, in distributed systems, it is impossible to isolate a single point, since each channel for the passage of energy is considered as a pair of poles (connection points) [Migulin et al. 1978], and therefore, the radiation frequency in the system will change in accordance with changes in the lengths of the combined fractures involved in the process of fault formation. This means that during the processes of closing and opening of cracks of different sizes in the focal zone and their ordering in a certain direction, which ultimately leads to the formation of the main fault, the emitted frequency in the system accurately reflects the change in the smallest distances between the pair of poles (connection points), which is also accurately reflected in the magnitude value.

The need for increased accuracy was clearly demonstrated by the example of an actual earthquake. In particular, in the case of hourly earthquake monitoring, it has become not only possible but also necessary to calculate changes in the length of a fault at the focus with an accuracy of up to a meter in the period preceding an earthquake [(Kachakhidze et al., 2024].

As mentioned above, the ultimate formation of the fracture length occurs by the rupture of a cofferdam between major fractures.

The analysis of the Crete (25.05.2016, M= 5.6) earthquake convinced us that the final formation of the fault length does indeed occur in accordance with the classical geological model. Before the first destruction of the cofferdams (16:00, 24 May), the length of the fault was 7828 meters. At the moment of the first destruction of the cofferdam, the length of the rupture became 7.5 km. That is, the length of the fault at the focus decreased by 328 m. It happened 16 hours before the earthquake. The process of the first destruction was expressed in the EM radiation in the form of an anomaly. The second destruction of another cofferdam occurred at 22:00 (24 May), which again was recorded by electromagnetic radiation anomaly. In result, the length of the rupture became 7.496 km. This time, the length of the fault decreased by 4 m. The earthquake occurred 16 hours and 36 minutes after the first anomaly and 10 hours and 36 minutes after the second, with the fault length decreasing by 21 m.

The above indicates that for a complete consideration of the complex geological process of earthquake preparation and occurrence, it is desirable to measure the fault length with an accuracy of meters by the formula (1).

Research on the Crete earthquake indicates, that on the scale of kilometers rupture occurring at the earthquake focus, is sensitive to the destruction on the scale of meters cofferdams [Kachakhidze at al., 2024].

Studies also confirm that during the nucleation stage, high-frequency FEMR (Fracture-Induced Electromagnetic Radiation) pulses in the MHz range appear as microcracks form. As the system transitions into the stick-slip phase, the frequency decreases to the kHz range [Rabinovitch et al., 2007, 2017]. Since the MHz range in the main corresponds to meter-scale ruptures, we believe they should be influenced by the destruction of centimeter-scale cofferdams.

Based on this view, it is better if the unified magnitude scale is divided into three subscales, depending on the accuracy of measuring the rupture length, where earthquakes are classified by magnitude as follows:

I. $\mathbf{1 \leq Ms \leq 3.909}$; II. $\mathbf{3.91 \leq Ms \leq 8.909}$ and III. $\mathbf{8.91 \leq Ms \leq 9}$

Furthermore, it should be noted that, as previously mentioned, electromagnetic radiation in the MHz range is observed even during the period preceding an earthquake [Eftaxias et al., 2002).

It is possible that, from a geological perspective, determining the length of the rupture in centimeters may not be of great importance; however, it is important for initial monitoring of the length of the rupture at the focus of an earthquake, even in the case of a strong earthquake, as well as for underground engineering applications. Thus, when determining length with such high precision using formulas (1), (2), and (3) for a specific electromagnetic radiation frequency, we obtain a precise and unique value for the magnitude and, consequently, a precise and unique value for the energy of the earthquake, which is of great importance in seismology.

### 2.3. Advantages and Potential Difficulties of EM Emissions That Existed Before Earthquakes

As it is known, for decades, theoretical, laboratory, and observational studies have been conducted on the EM radiation that occurs before an earthquake, which has revealed the advantages of this field compared to other geophysical field anomalies related to earthquakes:

1. Scientists note that electromagnetic radiation before an earthquake, at very large distances from the earthquake's epicenter, can be observed [Gokhberg et al., 1989; Gufeld et al., 1992, Hayakawa, 2007, Li et al, 2020]. For example, at the time of the earthquake in Crete, between the electromagnetic radiation receiver and the epicenter of the earthquake, the distance was approximately 200 km. [Kachakhidze et.al. 2021].

2. Scientific studies confirm that pre-earthquake electromagnetic emissions are observed considerably long before the earthquake [Biagi et al., 2013; Zhang, et al., 2011; Papadopoulos et al., 2010; Eftaxias et al., 2002, Li et al, 2020].

3. Our theoretical and retrospective material studies have revealed a key and decisive advantage in this field, that electromagnetic radiation in the range 23,830 kHz $\geq$ f $\geq$ 0.378 kHz is responsible for earthquakes with magnitude $1 \leq M \leq 9$.

4. Formula (1) is valid for earthquakes of any magnitude and can be applicable in all seismically active countries and regions, because, as it was mentioned above, in a system where earthquakes are being prepared (distributed system), it is impossible to isolate a single point, since each channel for the passage of energy is considered as a pair of poles (connection points) [Migulin et al. 1978], Therefore, the radiation frequency in the system will change in accordance with changes in the lengths of the combined fractures involved in the process of fault formation. This means that during the processes of closing and opening of cracks of different sizes in the focal zone and their ordering in a certain direction, which ultimately leads to the formation of the main fault, the emitted frequency in the system accurately reflects the change in the smallest distances between the pair of poles (connection points), which is also accurately reflected in the magnitude value.

Thus, the problems associated with the precise measurement of the magnitude of earthquakes of varying strengths, as well as the limitations of the applicability of the magnitude estimation formulas existing only for one region (country), are possible to be solved.

5. Based on the above, it is clear that the high accuracy of measuring the frequency of VLF/LF emissions (0.1 Hz) that existed before an earthquake allows, based on formula (1), to determine with very high (meter or centimeter) accuracy the length of the rupture occurring in the earthquake's focus and, consequently, allows measure the magnitude of an earthquake to within 0.001 for any region.This, in turn, in any region allows us to accurately estimate the energy of earthquakes, which is of crucial importance in modern seismology.

From the above, it is clear that the formula created by famous seismologists for any country or region, which links the length of the fault at the focus of an earthquake to the magnitude of the earthquake, is completely acceptable. It is also worth noting that the characteristic parameters of a future earthquake can be determined with such precision in advance, before it occurs.

6. Calculations for magnitudes Ms and Mw, based on the frequency of VLF/LF emissions that existed before the earthquake, accurately reflect the entire spectrum of earthquake magnitudes, from small to medium and large.

7. It is worth emphasizing that the VLF/LF electromagnetic radiation that existed before an earthquake always appears from the moment when the avalanche-like unstable process begins, and therefore, monitoring the frequency of this field and, hence, all other parameters becomes possible precisely from this moment on.This means that the length of the main fault of an incoming earthquake (i.e., magnitude), several tens of days before the earthquake can be determined.

8. Seismology currently lacks a reliable criterion for distinguishing strong foreshocks from the main shock. This question can be answered with a fairly high degree of accuracy using a theoretical model [Kachakhidze et al., 2015]: a trend toward decreasing electromagnetic radiation if is observed after any tremor, this indicates that fault formation at the earthquake's epicenter has not yet been completed, and the main shock should be expected. The situation is reversed with aftershocks.

9. Due to the determination in advance of fault length and magnitude with high accuracy, it also becomes possible to calculate in advance with high accuracy: the surface rupture length, downdip rupture width, rupture area, maximum and average displacement per event, the average subsurface displacement on the fault plane, seismic moment, and others [Wells et al., 1994].

10. Although the paper proposes a completely new parameter (VLF/LV electromagnetic radiation that existed before the earthquake) to determine the length of the rupture formed at the earthquake focus, it should be emphasized that this study does not take into account errors in formulas (2-5), which may be insignificant, but when monitoring real earthquakes can still appear.Therefore, from our point of view, it is desirable to specify the real errors when using formulas [2-5].

## 3. Conclusions

To obtain specific and unique values of earthquake characteristic magnitudes, the article proposes a completely new parameter, the frequency of the earthquake's prognostic precursor, i.e. the frequency of the VLF/LF electromagnetic emissions, that existed before the earthquake, and calculations based on it. This allows us:

1. During the monitoring of the earthquake preparation process, the frequency of VLF/LF electromagnetic radiation has to be measured within 0.1 Hz.

2. Based on the frequency of VLF/LF electromagnetic radiation, the length of the rupture occurring in the earthquake focus in advance, before the onset of an earthquake, has to be precisely determined with high accuracy (km, m, cm).

3. Based on the rupture length, the magnitude of a future earthquake has to be determined to within 0.001.

4. According to the one-thousandth magnitude, the energy of the upcoming earthquake should be measured (in joules).

5. Such precision measuring of these parameters was chosen because, in that case, one concret, unique value of the fault length corresponds to a concrete, unique values of the magnitude and energy of the earthquake.

6. The magnitude, fault length, and earthquake energy with high precision early detection, it becomes possible to resolve various geophysical and engineering problems in advance, including those related to seismic hazard.

We propose that various international agencies, such as the ISC, NEIC, and GCMT, consider adopting the method we have presented for measuring magnitude with a precision of one-thousandth, which is completely different from existing magnitude measurement methods.

It is based on a high-precision numerical estimate of the rupture length at the earthquake source, which, in turn, is determined from the frequency of electromagnetic emissions in the VLF/LF ranges observed before the earthquake. Measuring the magnitude with a precision of one-thousandth ensures the uniqueness of the earthquake's corresponding energy.

Author Contributions: B. K., converted every minute amplitudes of electromagnetic radiation data recorded by the INFREP network into frequency numerical values using the normal distribution of Gauss, G. R., and V.K. took part in the mathematical elaboration of data, prepared INFREP network VLF/LF electromagnetic radiation data for processing, M. K., and N. K-M., set the tasks, equally took part in the mathematical elaboration of data, analyzed results, wrote and revised the manuscript.

Conflicts of Interest
The authors declare no conflict of interest.

**Acknowledgments**

The authors are grateful to the network INFREP for providing us with the MHz and kHz fractal-electromagnetic emissions data used in this paper.